\documentclass[aps,preprint]{revtex4}
\usepackage{graphicx}

\begin{document}

\title{Spanning Trees in Random Satisfiability Problems}

\author{A. Ramezanpour} \email{ramezanpour@iasbs.ac.ir}

 \affiliation{Institute for Advanced Studies in Basic Sciences,
Zanjan 45195-1159, Iran}

\author{S. Moghimi-Araghi}
\email{samanimi@sharif.edu}

\affiliation{Department of Physics, Sharif University of
Technology, P.O.Box 11365-9161, Tehran, Iran}

\date{\today}

\begin{abstract}
Working with tree graphs is always easier than with loopy ones and
spanning trees are the closest tree-like structures to a given
graph. We find a correspondence between the solutions of random
K-satisfiability problem and those of spanning trees in the
associated factor graph. We introduce a modified survey
propagation algorithm which returns null edges of the factor graph
and helps us to find satisfiable spanning trees. This allows us to
study organization of satisfiable spanning trees in the space
spanned by spanning trees.
\end{abstract}

 \maketitle

\section{Introduction}\label{1}
Understanding the nature of complexity in NP-Hard optimization
problems is an important goal in the study of complex systems
\cite{ks,mzkst}. Most of optimization problems are indeed a
constraint satisfaction problem and satisfiability problem is a
prototype of such problems \cite{m1,mpz}. In a satisfiability
problem one looks for a configuration of logical variables that
satisfies a number of logical constraints (clauses). A clause is in turn
disjunction of a few logical variables. In random
K-satisfiability (K-SAT) problem a clause involves $K$ randomly
selected variables each negated with a given probability.
Increasing the number of clauses per variables, these problems
display a phase transition from satisfiable to unsatisfiable phase
\cite{ks,mz0}. It is around this SAT-UNSAT transition that the
typical complexity of the problem raises rapidly. In fact the
transition is preceded by a clustering of solutions in
the configuration space, a phenomenon known as replica symmetry
breaking in physics literatures \cite{m2,mmz}.
There are a few analytic methods and algorithms, such as replica
and cavity methods and survey propagation algorithm
\cite{mz,mez,bmz}, that deal with the behavior of random
satisfiability problems.\\
Here we are going to look at the problem from a rather geometrical
point of view. We know that there are efficient algorithms such as
WalkSAT and Belief propagation \cite{bhw,kfl,bmz} that can
efficiently solve a random K-SAT problem possessing a tree factor
graph; A factor graph is a bipartite-graph representation of
constraint satisfaction problems. We also know that spanning trees
are the closest tree structures to a given graph. Then it is
interesting to find the relation between the solutions of a factor
graph and the solutions of its spanning trees. This relation could
result in
the reduction of the original problem to easier ones.\\
In this paper we show that there is indeed a correspondence
between the solutions of a random K-SAT problem and those of
spanning trees in the associated factor graph. In principle to
check that a random K-SAT problem is satisfied or not one has to
check the satisfiability of all the spanning trees. The main
problem is that in general the number of spanning trees grows
exponentially with the size of graph. Moreover we will see that
even when we are in the SAT phase almost all randomly generated
spanning trees are unsatisfiable. To get around this problem we
assign a weight to each edge of the factor graph. For an edge, its
weight is taken as the probability that the edge to be present in
the graph. Then we suggest an ansatz that relates the above
probability to the survey of warnings sent along the edge and run
survey propagation algorithm. This algorithm converges with a nonzero probability in
the SAT phase and we find that in this case all maximum spanning
trees (which have the maximum sum of weights on their edges) are
satisfied. In other words, the above algorithm defines null edges
of the factor graph.\\

The structure of paper is as follows; In section \ref{2} we define
the problem more precisely and illustrate the relation between a
factor graph and its spanning trees. We briefly deal with the
satisfiability of randomly generated spanning trees in section
\ref{3}. In section \ref{4} we modify the survey propagation
algorithm to generate satisfiable spanning trees. Section \ref{5}
includes our concluding remarks.

\section{From factor graph to its spanning trees}\label{2}
A random satisfiability problem is defined as follows: We take $N$
logical variables $x_i\in \{0,1\}$. Then we construct a formula
$F$ of $M$ clauses joined to each other by logical AND. Each
clause contains a number of randomly selected logical variables.
In random K-SAT problem each clause has a fixed number of $K$
variables. These variables, which join to each other by logical
OR, are negated with probability $p$ and appear as such with
probability $1-p$. For example $F:=( \overline{x}_2\vee
x_4)\wedge(\overline{x}_3\vee x_2)\wedge(x_1 \vee x_3)$ is a 2-SAT
formula with $3$ clauses and $4$ logical variables. A solution of
$F$ is
a configuration of logical variables that satisfy all clauses.\\
The relevant parameter that determines the satisfiability of $F$
is $\alpha:=M/N$. In the thermodynamic limit ($N,M \rightarrow
\infty$ and $\alpha \rightarrow const.$) $F$ is satisfied with
probability one as long as $\alpha < \alpha_c$. Moreover, it has
been found that for $\alpha_d<\alpha<\alpha_c$ the problem is in
the Hard-SAT phase \cite{mez}. At $\alpha_d$ we have a dynamical
phase transition associated with the break down of replica
symmetry. Assuming one-step replica symmetry breaking, one obtains
$\alpha_d \simeq 3.92$ and $\alpha_c\simeq4.26$ for random 3-SAT
problems with $p=1/2$ \cite{mez}. Although this approximation
seems to be exact near the SAT-UNSAT transition but it fails close
to the dynamical transition where higher order replica symmetry
breaking solutions are to be used \cite{mtb,smz}. The above
approximation allows us to compute the configurational entropy or
complexity of the problem. The complexity, which is logarithm of
the number of clusters in the solution space,
takes abruptly a nonzero value at $\alpha_d$ and decreases to zero at $\alpha_c$.\\
The factor graph is a bipartite graph of variable nodes and
function nodes (clauses). The structure of this graph is
completely determined by an $M\times N$ matrix with elements
$J_{a,i}\in \{0,+1,-1\}$; $J_{a,i}=+1$ if clause $a$ contains
$x_i$, it is equal to $-1$ if $\overline{x}_i$ appears in $a$ and
otherwise $J_{a,i}=0$.\\
Now consider a random K-SAT problem and the associated factor
graph. In the SAT phase the problem has some solutions that
satisfy all the clauses. The factor graph has a number of spanning
trees which have their own solutions. These solutions can be
easily obtained by Walksat algorithm or some other local search
algorithms \cite{sao}. Notice that a spanning tree could have no
solution even when the original problem is satisfied. Here we show
that (i) \textit{Each solution of a spanning tree is the solution
of the original graph too.} And (ii) \textit{Each solution of the
original graph is the solution of at
least one of the spanning trees}.\\
Note that in a spanning tree
each function node is connected to a smaller number of variable
nodes than in the original graph. Thus if a configuration of
variable nodes satisfies a spanning tree it should also satisfy
the original problem.\\
To verify the second statement we start with a solution of the
problem and try to construct a spanning tree that is also satisfied by
this solution. To this end we have to remove some of the edges in
the graph until no cycle remains in the factor graph. During this
process we should, of course, respect the graph connectedness and
more importantly satisfaction of all the function nodes. Therefore,
at each step we consider a cycle of the graph and
remove one of the null edges along this cycle. A null edge is an
edge that its removal dos not change the number of satisfied
clauses. This removal also preserves the graph
connectedness. Note that a null edge at a given stage may not be a
null one in the next stages.\\ To complete the proof we should
show that for a solution of the factor graph, any cycle in the graph
has at least one null edge. First notice that each function node
which lies on a cycle has at least two neighbors. Since we are
dealing with a solution then at least one of its neighboring
variable nodes satisfies the function node. In this case at least
one of the edges along the cycle will be a null edge and we can
freely remove it from the graph.\\
Indeed for any solution, a cycle would have a large number of null
edges which result in a large degree of freedom in constructing
the desired spanning tree. The number of null edges along a cycle
would be of order of the number of function nodes in the cycle. It
means that an arbitrary solution appears in a large number of
spanning trees.\\
Suppose that we have a satisfiable spanning tree. Then we can use
a local search algorithm to find its solutions. Notice that for a tree
factor graph there is no frustration in the problem and one can safely
assume all the solutions in the same cluster. Indeed, Warning propagation
algorithm \cite{kfl} which is based on this assumption gives exact results
for tree factor graphs.  Thus we expect that the solutions of a satisfiable
spanning tree lay in the same cluster of solutions. In the thermodynamic limit, it
means that for any two solutions, there is at least one path in
the solution space that connects them to each other. The path
starts from one of the solutions and reach the other one by
successive steps of finite lengths. Since the solutions also satisfy the original problem
we may conclude that they lay in the same cluster of the original problem too.\\
Please notice that one could use the same arguments (as given in
this section) to show that any configuration of variables that
violates $E$ clauses in the original factor graph also appears in
at least one of the spanning trees. Therefore, the state with
minimum $E$ (the ground state) in the original factor graph, has
also the minimum $E$ among the ground states of all the spanning
trees.

\section{Satisfiability of spanning trees}\label{3}
There is an algorithm that uses a simple random walker on the
original factor graph and generates (with a uniform measure) a
spanning tree \cite{b}. Using this algorithm we generated up to
$10^7$ spanning trees of the factor graph of a random 3-SAT
problem with $N=100$. Belief propagation algorithm can be used to
compute the entropy (logarithm of the number of solutions) of each
spanning tree. We found that when $p=0$ both the average entropy
and its dispersion decrease exponentially with $\alpha$. However,
if we increase $p$ slightly, we find that almost all the randomly
generated spanning trees have no solution. For instance, with
$\alpha=4$ and $p=0.1$, there is almost no satisfiable spanning
tree among $10^7$ randomly generated ones. In this case, in
average about $25$ clauses are not satisfied. When we increase $p$
to $0.5$ the average number of unsatisfied clauses in a randomly
generated spanning tree increases to $100$ whereas the original
problem is satisfiable. We found that any naive attempt, like the
generalization of simple random walk procedure to a biased one,
fails in generating satisfiable spanning trees. These observations
show that finding satisfiable spanning trees is not an easy task.
In the following we look for a more sophisticated way to solve
this problem. We will restrict ourselves to random 3-SAT problems
in the most random case, i.e. $p=1/2$.
\section{Generating satisfiable spanning trees}\label{4}
\begin{figure}
\includegraphics[width=8cm]{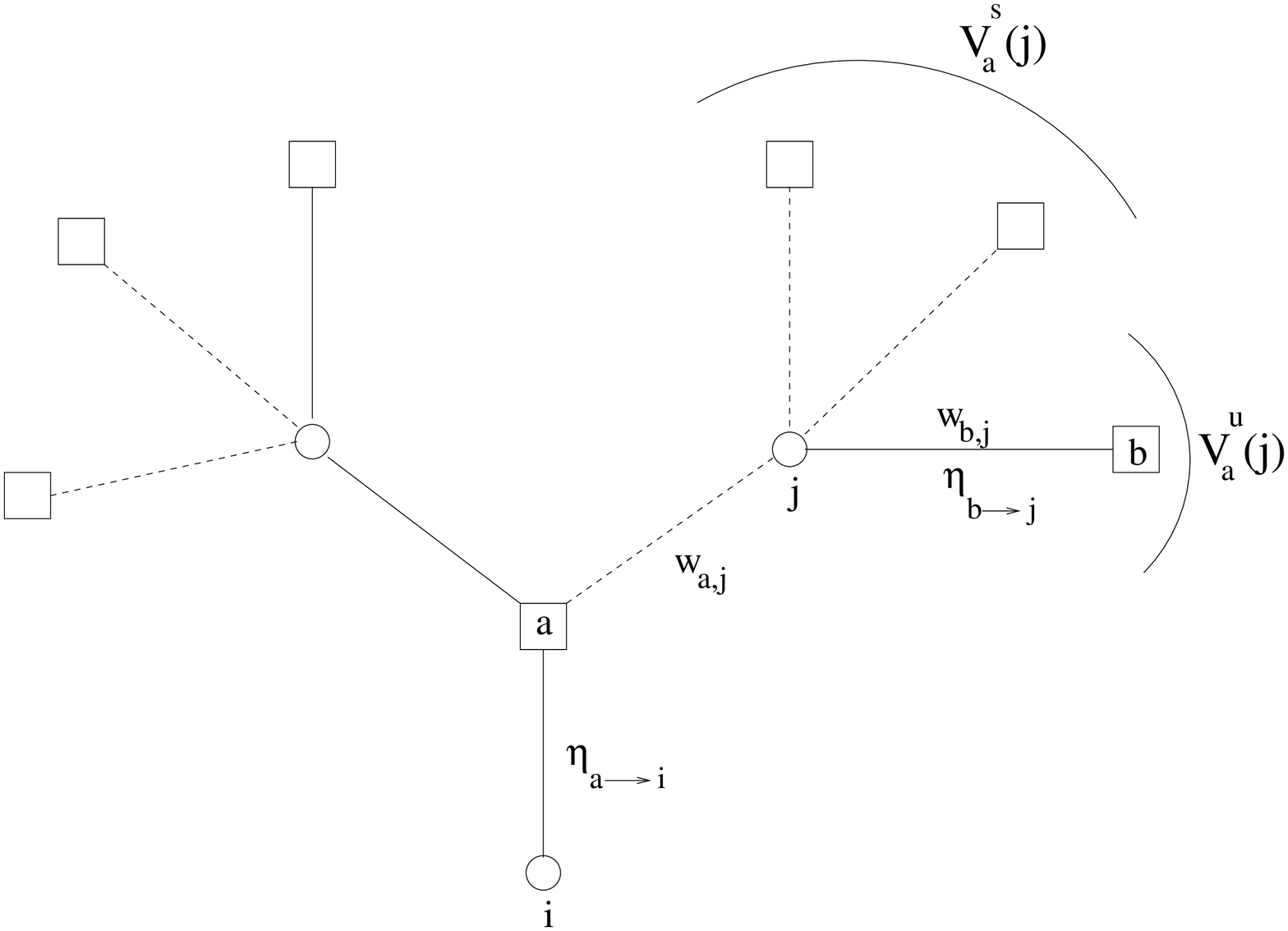}
\caption{The survey $\eta_{a\rightarrow i}$ is determined by the
set of surveys $\eta_{b\rightarrow j}$ and weights $w_{a,j},
w_{b,j}$. In this factor graph squares and circles represent
clauses and variables respectively. A dashed line shows that the
negated variable appears in the clause.}\label{f1}
\end{figure}

\begin{figure}
\includegraphics[width=8cm]{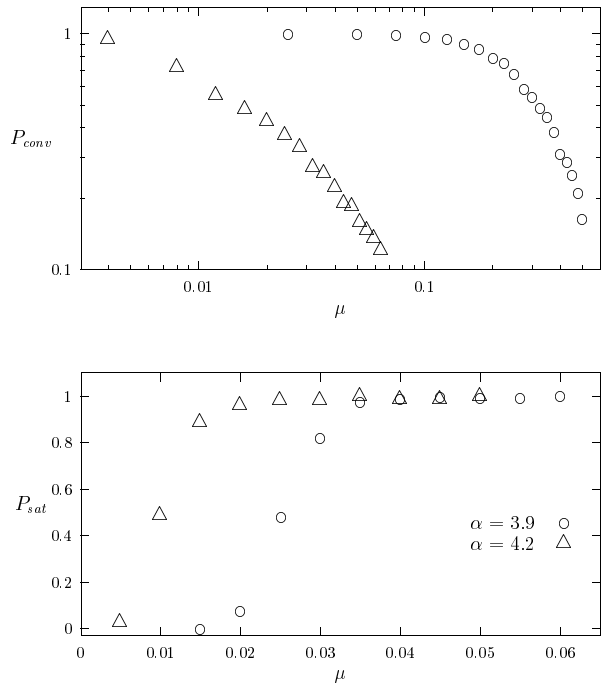}
\caption{ Convergence probability ($P_{conv}$) and satisfiability
probability of a maximum spanning tree ($P_{sat}$) versus $\mu$
for a random 3-SAT problem with $N=1000$ and $p=1/2$. The data for
$P_{conv}$ obtained from $2000$ realizations of the problem. To
obtain $P_{sat}$ we randomly generated $20$ maximum spanning trees
in $150$ satisfiable realizations. }\label{f2}
\end{figure}

\begin{figure}
\includegraphics[width=8cm]{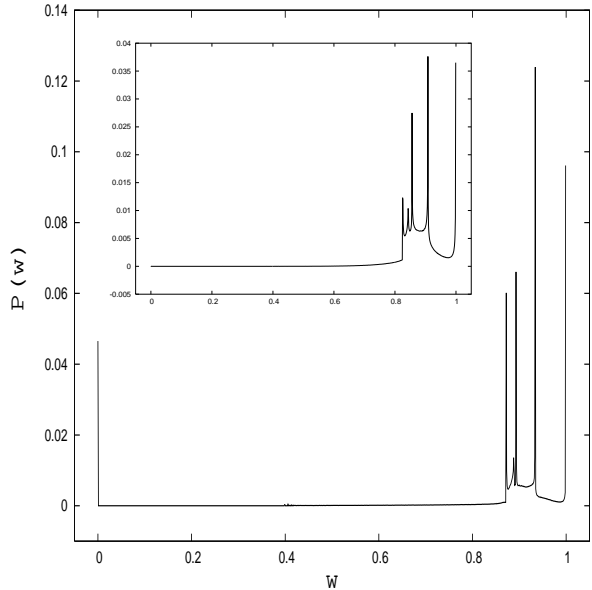}
\caption{ Weight distribution of edges in a random 3-SAT problem
with $N=1000$, $p=1/2$ and $\alpha=4.2, 3.9 (inset)$ when $\mu$ is
$0.025,0.04$ respectively. The data are result of averaging over
$1000$ realizations }\label{f4}
\end{figure}

\begin{figure}
\includegraphics[width=8cm]{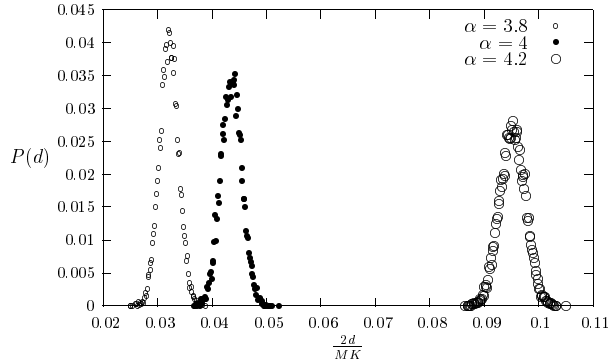}
\caption{$P(d)$ for a single realization of random 3-SAT problem
with $N=1000$ and $p=1/2$. For $\alpha=3.8, 4, 4.2$ ($\mu=0.055,
0.035, 0.025$ respectively) we have generated $10^4$ pairs of
satisfiable spanning trees. }\label{f5}
\end{figure}

In section \ref{2} we used the notion of null edges to construct
spanning trees that are satisfied by a given solution. It suggests
that some of the edges can be removed from the factor graph with
no effect on the number of satisfied clauses. But how can we
distinguish these null edges? Suppose that we remove some edges
and run the survey propagation algorithm \cite{bmz}. If the
problem is in the SAT phase the algorithm converges with a nonzero
probability and for each edge returns a survey of warnings
$\eta_{a\rightarrow i}$. The survey $\eta_{a\rightarrow i}$ is the
probability that in an arbitrary cluster of solutions, clause $a$
sends a warning to variable $i$. This warning enforces $i$ to take
a value which satisfies $a$. If the problem is in the UNSAT phase
then the algorithm may still converge but it results to a negative
value for the complexity. In the following we will work with a
version of survey propagation algorithm in which we assign zero
probability to contradictory massages. Then the algorithm can end with
a set of contradictory surveys. In this case a variable receives surveys
that suggest different values for it and taking any value results in at
least one unsatisfied clause.\\
Let us define  $w_{a,i}$ as the probability that edge $(a,i)$ to
be present in the factor graph. Then we run the survey propagation
algorithm taking these probabilities into account. It means that
when function node $a$ sends a message to its variable node $i$,
it gives more weights to the neighbors having a higher probability
of being present in the graph. More precisely, $\eta_{a\rightarrow
i}$ is given by
\begin{equation}\label{eta0}
\eta_{a\rightarrow i}=\prod_{j\in
V(a)-i}[w_{a,j}P_a^u(j)+1-w_{a,j}],
\end{equation}
where $P_a^u(j)$ is the probability that variable $j$ can not
satisfy clause $a$. We also denote by $V(a)$ the set of variables
belong to clause $a$ and by $V(i)$ the set of clauses that
variable $i$ contributes in. In survey propagation algorithm $P_a^u(j)$ is given by \cite{bmz}
\begin{equation}\label{Pauj}
P_a^u(j)=\frac{\Pi_{j\rightarrow a}^u}{\Pi_{j\rightarrow
a}^s+\Pi_{j\rightarrow a}^0+\Pi_{j\rightarrow a}^u},
\end{equation}
where
\begin{eqnarray}\label{Pii}
\Pi_{j\rightarrow a}^0=\prod_{b\in
V(j)-a}\left(1-w_{b,j}\eta_{b\rightarrow j}\right), \\
\nonumber \Pi_{j\rightarrow a}^u=[1-\prod_{b\in
V_a^u(j)}\left(1-w_{b,j}\eta_{b\rightarrow j}\right)]
\prod_{b\in V_a^s(j)}\left(1-w_{b,j}\eta_{b\rightarrow j}\right),\\
\nonumber \Pi_{j\rightarrow a}^s=[1-\prod_{b\in
V_a^s(j)}\left(1-w_{b,j}\eta_{b\rightarrow j}\right)]\prod_{b\in
V_a^u(j)}\left(1-w_{b,j}\eta_{b\rightarrow j}\right).
\end{eqnarray}
Here $V_a^s(j)$ denotes to the set of clauses in $V(j)-a$ that
variable $j$ appears in them as it appears in clause $a$, see Fig.
\ref{f1}.
The remaining set of clauses are denoted by $V_a^u(j)$.
We should also give a way to determine $w_{a,i}$. Note that we
expect a positive correlation between $w_{a,i}$ and
$\eta_{a\rightarrow i}$. If the survey is large, the edge is
necessary for satisfaction of clause $a$ and in the opposite case
its presence is not essential. Our suggested ansatz for the
weights is
\begin{equation}\label{weta}
w_{a,i}=[\eta_{a\rightarrow i}]^{\mu},
\end{equation}
where $\mu \ge 0$ controls the number of removed edges. For each
instance of the problem we start by a random initialization for
$\eta$'s and $w$'s. In each iteration we select all of the edges
in a random way. When an edge is selected we first update
$\eta_{a\rightarrow i}$ and then $w_{a,i}$ according to Eqs.
(\ref{eta0}-\ref{weta}). We expect that for appropriate values of
$\mu$ the above algorithm converges (without contradictory
surveys) as long as we are in the SAT phase. In practice we define
a maximum $t_{max}$ for the number of iterations $t$. Thus for a
given $\mu$ it is possible not to reach a convergence in $t<
t_{max}$. We define $P_{conv}$ as the probability that a randomly
generated instance of the problem converges in a number of
iterations less than $t_{max}$. In our numerical simulations we
will take $t_{max}=1000$ and the limit of convergence is taken
$\epsilon=10^{-3}$. Note that when $\mu=0$ we recover the usual
survey propagation algorithm on the original factor graph. In this
situation the algorithm always converges and $P_{conv}=1$.
Increasing $\mu$ means that we are removing some of the edges and
so we expect more time to reach a convergence. Thus
$P_{conv}(\mu)$ would be a decreasing function of its argument. We
find that its behavior depends on $\alpha$, Fig. \ref{f2}. In
fact, $P_{conv}$ diminishes more rapidly for
greater values of $\alpha$.\\
Suppose that our algorithm converges. Then we expect that a
spanning tree with maximum sum of weights for its edges, i.e. a
maximum spanning tree, be a satisfiable one. When the algorithm
converges and returns the $w$s we can construct a maximum spanning
tree as follows: Starting from a randomly selected node in the
original factor graph we first find the maximum weight among the
edges that connect it to the other nodes. Then we list the edges
having a weight in the $\epsilon$-neighborhood of the maximum one
and add randomly one of them to the new factor graph. If we repeat
the addition of edges $N+M-1$ times we obtain a spanning tree
factor graph which is a maximum spanning tree. Notice that taking
a nonzero interval to define the edges of maximum weight at each
step, along with the randomness in choosing one of them, allow to
construct a large number of maximum spanning trees. Let us define
$P_{sat}(\mu)$ as the probability that a maximum spanning tree be
satisfiable if the algorithm converges. The best value of $\mu$ is
one that for which $P_{sat}$ takes its maximum value. In Fig.
\ref{f2} we also show $P_{sat}(\mu)$ for two values of
$\alpha=3.9,4.2$. As the figure shows $P_{sat}\simeq 1$ when $\mu
> 0.04$  ($\alpha=3.9$) and $\mu > 0.025$ ($\alpha=4.2$).
However, as stated above, by increasing $\mu$ the convergence time
increases. Thus the optimal $\mu$ for $\alpha=3.9,4.2$ will be
around $0.04$ and $0.025$ respectively. The optimal $\mu$
decreases with $\alpha$. It means that as we get closer to the
SAT-UNSAT transition we have to give more
importance to the surveys.\\
\begin{figure}
\includegraphics[width=8cm]{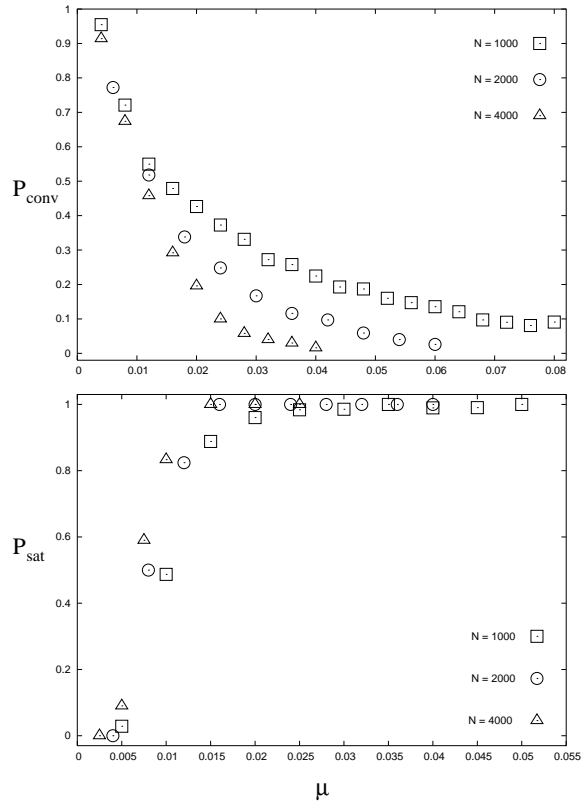}
\caption{ Convergence probability ($P_{conv}$) and satisfiability
probability of a maximum spanning tree ($P_{sat}$) versus $\mu$
for random 3-SAT problems of different sizes with $p=1/2$. Here the statistical
errors are of order $0.01$.}\label{f3}
\end{figure}
To see the finite size effects we run the algorithm for larger problem sizes
$N=2000,4000$. The results for $P_{conv}$ and $P_{sat}$ at $\alpha=4.2$ have been
shown in Fig. \ref{f3}. The convergence probability decreases more rapidly with
$\mu$ as $N$ increases. That is we have to pay more computational efforts to
get a converged situation. On the other side, by increasing $N$ the
satisfiability probability enhances for smaller values of $\mu$ and
gets more rapidly its saturation value. We hope that this behavior
of $P_{sat}$ compensate the decrease in $P_{conv}$ for larger problem sizes.
As Fig. \ref{f3} shows the changes in the quantities are significant and
we have a strong size dependence for the small values of $N$ studied here.\\
Finally let us look at the statistics of $w_{a,i}$'s and the
organization of maximum spanning trees in the space spanned by
spanning trees.\\ We observe that if $P_{sat}\simeq 1$ then the
weight distribution of the edges, $P(w)$, is almost sum of a few
delta peaks.  Fig. \ref{f4} displays this behavior of $P(w)$ for
$\alpha=3.9,4.2$. In both cases, edges of weight $1$ have a
considerable contribution and the weight distribution dose not
define a single characteristic weight. We think that it is these
properties of $P(w)$ that allow the maximum-spanning-tree scenario
to work. A clear difference between the two distributions in Fig.
\ref{f4} is the presence of a delta peak at $w=0$ for
$\alpha=4.2$. We found that this peak appears continuously around
$\alpha \simeq 3.92$, where we expect the dynamical phase transition to happen.\\
We define the distance between two spanning trees as the number of
uncommon edges. Then we take two randomly generated satisfiable
spanning trees and obtain their distance $d$. We repeat this for a
number of times and construct $P(d)$, the normalized number of
times that the two spanning trees are at distance $d$ of each
other. The result has been shown in Fig. \ref{f5} for $\alpha=3.8,
4, 4.2$. In these cases $P(d)$ is a Gaussian distribution and
defines a characteristic distance which grows with $\alpha$.
Moreover, as the figure displays, by increasing $\alpha$
dispersion around the characteristic distance increases. Note that
more distant factor graphs are more likely to have distant
solutions. Therefore, we may conclude that as satisfiable spanning
trees get away from each other the same happens in the solution
space. These facts are consistent with the present picture
\cite{m1,smz} of the clustering phenomenon in the solution space.
However, the presented data in Fig. \ref{f5} are not sufficient to
detect the clustering phase transition in the solution space. To
this end we should, probably, study the behavior of the average
distance and its dispersion for different problem sizes.

\section{Conclusion}\label{5}
We showed that there is a correspondence between the solutions of
a random K-SAT problem and those of spanning trees in the
associated factor graph. This relation may be useful in finding
more rigorous results for the structure of the solution space.
The concept of null edges helped us to modify survey propagation
algorithm and to construct satisfiable spanning trees. A
satisfiable spanning tree has a number of solutions (in the same
cluster of solutions) that can easily be obtained by local search
algorithms. This provides us another way of extracting solutions
of a random K-SAT problem. We used the modified algorithm to study
organization of satisfiable spanning trees in the space of
spanning trees. We found that by increasing the number of clauses
per variable, the distance between satisfiable spanning trees
increases. This finding supports the previous results on the clustering
in the solution space.\\
In this paper we focused on the case $K=3$ and $p=1/2$ but we
found that the algorithm works well for $K=4,5$ and other values
of $p$ (not represented here).

\acknowledgments We would like to thank M. Mezard that encouraged
us to study this problem.

\end{document}